# Optical Kerr soliton crystal microcomb source for RF photonic fractional differentiation

Mengxi Tan, Xingyuan Xu, Bill Corcoran, Jiayang Wu, Andreas Boes, *Member, IEEE*, Thach G. Nguyen, Sai T. Chu, Brent E. Little, Roberto Morandotti, *Senior Member, IEEE*, Arnan Mitchell, *Member, IEEE*, and David J. Moss, *Fellow, IEEE*

*Abstract*—We report a photonic radio frequency (RF) fractional differentiator based on an integrated Kerr micro-comb source. The micro-comb source has a free spectral range (FSR) of 49 GHz, generating a large number of comb lines that serve as a high-performance multi-wavelength source for the differentiator. By programming and shaping the comb lines according to calculated tap weights, arbitrary fractional orders ranging from 0.15 to 0.90 are achieved over a broad RF operation bandwidth of 15.49 GHz. We experimentally characterize the frequency-domain RF amplitude and phase responses as well as the temporal responses with a Gaussian pulse input. The experimental results show good agreement with theory, confirming the effectiveness of our approach towards high-performance fractional differentiators featuring broad processing bandwidth, high reconfigurability, and potentially greatly reduced size and cost.

## I. INTRODUCTION

The exploding demand for processing speed and throughput in modern electronic systems has spurred a renewed interest in analog radio-frequency (RF) components. Photonic RF techniques, which perform signal processing in the optical domain via electrical-to-optical (EO) or optical-to-electrical (OE) conversion, offer a number of advantages over their electronic counterparts [1]. They have much higher speed and bandwidth, in some cases orders of magnitude higher than electronic signal processors, and thus have wide applications ranging from antenna remoting in radar systems, satellite communications, and cable television signal distribution [2-4].

As a key basic function in RF signal processing systems, differentiation has wide applications in ultra-wideband (UWB) generation, RF spectrum analysis, and RF filters [5-9]. Significant progress has been made on RF integral differentiators using photonics [10-22], such as using phase [10, 11] and cross-phase modulation (XPM) [15-17] methods, frequency discriminators and cross-gain modulation in semiconductor optical amplifiers (SOA) [14].

Fractional differentiation is a more generalized version of integral differentiation and has arguably many more unique and powerful applications [20, 21]. It has played an important role in many physical sciences such as mechanics, electricity, chemistry, biology, and economics. Perhaps its most important applications have come in the realms of image edge detection, control theory, and mechatronics [20, 21]. Despite this, however, photonic based fractional differentiators have experienced comparatively little attention.

Many photonic based approaches for both integral and non-integral differentiation have focused on producing the derivative of the complex optical field, rather than the pure RF differentiation. A photonic differentiator based on a dual-drive Mach-Zehnder modulator together with an RF delay line was recently reported [12] that, although successful, was intrinsically limited in processing speed by the operational bandwidth of the RF delay line. RF differentiators based on optical filters (OFs) have also been reported [13] that feature high processing speeds of up to 40-Gb/s, although this approach works only for a fixed (and typically integral) differentiation order and lacks reconfigurability.

To implement highly reconfigurable differentiators, transversal schemes have been investigated using discrete laser arrays [7, 18, 19], although at the expense of significantly increased system size, cost, and complexity, limiting the number of available taps and thus the processing performance. Therefore, instead of using individual light sources for each tap, a single source that can simultaneously generate a large number of high-quality wavelength channels would be highly advantageous.

Integrated micro-combs [22-34], generated through optical parametric oscillation in monolithic micro-ring resonators (MRRs), are promising multi-wavelength sources for RF signal processing [35-44] that offer many advantages including a much higher number of wavelengths and greatly reduced footprint and complexity. Recently [35], we demonstrated a photonic RF differentiator with 8 taps using a 200 GHz spaced micro-comb source and achieved the first-, second-, and third-order differentiation functions operating at an RF bandwidth of up to 17 GHz with a potential bandwidth up to 100 GHz.

In this paper, we demonstrate the first fractional-order photonic based RF differentiator that operates directly on the RF signal, rather than on the complex optical field. It is based on a 49 GHz spaced Kerr micro-comb source, which provides a large number of comb lines (up to 80 in the telecommunications C band) that in turn allow a broad operation bandwidth of 15.49 GHz. By programming and shaping the comb lines according to calculated tap weights, we experimentally demonstrate reconfigurable arbitrary fractional orders ranging from 0.15 to 0.9, for which we characterize the RF amplitude and phase response. System demonstrations of real-time fractional differentiation of Gaussian input pulses are also performed. The good agreement between theory and experiment confirms our approach as an effective way to implement high-speed reconfigurable fractional differentiators with reduced footprint, lower complexity, and potentially lower cost.

The fractional differentiator is based on a novel and powerful form of microcomb called "soliton crystals" [30], realized in a CMOS compatible platform [23, 24]. Combined with the micro-comb's low FSR spacing of 48.9 GHz this results in potentially 80 wavelengths, or taps, being generated in the telecommunications C-band. Our results stem from the soliton crystal's extremely robust and stable operation and generation as well as its much higher intrinsic efficiency, all of which are enabled by an integrated CMOS-compatible platform.

## II. PRINCIPLE OF OPERATION

An $N_{th}$-order temporal RF fractional differentiator can be considered as a linear time-invariant system with a transfer function given by [1]

$$H_N(\omega) \propto (j\omega)^N \qquad (1)$$

where $j = \sqrt{-1}$, $\omega$ is the RF angular frequency, and $N$ is the differentiation order which can be fractional and even complex [20, 21]. According to the above transfer function, the amplitude response of a temporal differentiator is proportional to $|\omega|^N$, while the phase response has a linear profile, with a phase jump of $N\pi$ at null frequencies.

In this paper, we employ the reconfigurable transversal approach to achieve a photonic RF fractional differentiator, where a finite set of delayed and weighted replicas of the input RF signal are produced in the optical domain and combined upon detection. The transfer function of a general transversal filter can be described as

$$F(\omega) = \sum_{k=0}^{M-1} h_k e^{-j\omega KT} \qquad (2)$$

where $M$ is the number of taps, $T$ is the time delay between adjacent taps, and $h_k$ is the tap coefficient of the $k_{th}$ tap, which is the discrete impulse response of $H_N(\omega)$. The discrete impulse response can be calculated by performing the inverse Fourier transform of $H_N(\omega)$, and then temporally windowed with a short cosine bell [45].

## III. EXPERIMENTAL RESULTS

The experimental setup of the fractional differentiator based on a Kerr micro-comb is shown in Fig. 1. We used an integrated multi-wavelength source—the microcomb, to generate the wavelengths that form the taps for the transversal structure. The soliton crystal microcomb was generated via on-chip optical parametric oscillation in a micro-ring resonator (MRR), which was fabricated with CMOS-compatible fabrications processes [26, 42] including doped silica glass deposition, patterning, and silica deposition. Due to the ultra-low loss of our platform, the MRR features narrow resonance linewidths of ~130 MHz [46, 47], corresponding to a quality factor of ~1.5 million. The waveguide dimension was tailored to have anomalous dispersion in the C band. The MRR was designed to have a radius of ~592 μm, corresponding to an optical free spectral range of ~0.4 nm or 48.9 GHz.

A continuous-wave pump at ~1550 nm was amplified by an optical amplifier, with the polarization state aligned with the TE mode of the MRR. The pump wavelength was swept manually until the proper detuning between the pump wavelength and the MRR's resonance was achieved to generate coherent soliton crystal states, which are characterized by the distinctive 'fingerprint' optical spectrum (Fig. 2) — a result of the spectral interference between tightly packaged solitons in the cavity [28].

The soliton crystal comb was then spectrally shaped with two commercial optical spectral shapers (Finisar, WaveShaper 4000S) to enable a better signal-to-noise ratio (with respect to a single Waveshaper) as well as a higher shaping accuracy. The first WaveShaper (WS1) flattened the microcomb and reduced the initial power difference between the comb lines to < 2 dB, while the second WaveShaper (WS2) accurately shaped the comb power according to the designed tap coefficients. A feedback control path was adopted to increase the accuracy of the comb shaping, where the shaped comb spectra from the WaveShapers were detected by an optical spectrum analyzer and compared with the ideal tap weights to generate feedback error signals for calibration. In the experiment, we employed every second comb line, and so the spectral spacing between adjacent wavelength channels was $\Delta\lambda$ = 0.8 nm.

Next, the shaped comb lines were modulated by the RF input signal, which was multicast onto the wavelength channels to yield replicas. These were then transmitted through a spool of standard single mode fibre ($L$ = 2.1 km, $\beta$ = ~17.4 ps / nm / km) to obtain a progressive time delay between adjacent wavelengths of $\tau = L \times \beta \times \Delta\lambda$ = 29.4 ps. Finally, the delayed replicas were combined and then summed upon photo-detection. Both positive and negative taps were achieved by separating the wavelength channels according to the symbol of the designed tap coefficients and then fed to a balanced photodetector (Finisar, BPDV2150R). By tailoring the comb lines' power according to the tap coefficients, arbitrary fractional orders can be achieved. The Nyquist bandwidth of the fractional differentiator was inversely related to the progressive delay of the taps $\tau$, which in this case was $1/2\tau$ = ~17 GHz. In principle this can be made much larger by using large FSR microcombs, up to 200 GHz in spacing [35].

A key advantage of our approach is the large number of taps brought about by the microcomb, as such, we first investigated the relationship between them and the performance of the fractional differentiator. Fig. 3 shows the transfer function of the fractional differentiator for six different orders. As can be seen, the operational bandwidth of the differentiator—within which the slope coefficients of the simulated and ideal amplitude responses match closely—increases with the number of taps. Fig. 4 (a) illustrates the further extracted relationships, showing that the number of taps is critical to the operation bandwidth of fractional differentiators, which is also manifested experimentally in the case of a fractional order = 0.45 (Fig. 4(b, c)). When the number of taps is increased to 27, the operational bandwidth of the fractional differentiator could reach 15.49 GHz, occupying > 91% of the Nyquist band. In the experiment, up to 29 combs lines (capable of offering a sufficiently large operation bandwidth) were employed to generate the taps. We note that to further increase the number of taps, the optical signal-to-noise ratio, which was subject to the optical amplifiers' noise in our case, should be optimized to > 40 dB.

The shaped comb spectra are shown in Fig. 5. The wavelength channels for positive and negative taps were separately measured by an optical spectrum analyser. The optical power for each comb line matched closely with the designed tap coefficients, verifying the success of our comb shaping procedure. The transmission response of the fractional differentiator was characterized using a vector network analyser (Agilent MS4644B). The measured |S21| curves denoted the power responses of the differentiator, where the operation bandwidth ranged from DC to 15.49 GHz. As shown in Fig. 6, the range of the fractional orders was associated with the close match between the experimental and simulated results in terms of the slope coefficients associated with the power responses and the phase shift in the measured phase responses.

We further verified the performance of the fractional differentiator using a baseband RF Gaussian pulse. The Gaussian pulse was generated by an arbitrary waveform generator (AWG, KEYSIGHT M9505A), featuring a full-width at half maximum (FWHM) of ~200 ps, as shown in Fig. 7. The differentiated waveforms were recorded by a high-speed real-time oscilloscope (KEYSIGHT DSOZ504A Infinium) and closely matched with their theoretical counterparts, indicating the range in orders that are achievable with our approach to the reconfigurable fractional differentiator.

## IV. Conclusion

We propose and demonstrate a photonic RF fractional differentiator based on an integrated Kerr micro-comb source that operates on the RF signal rather than the complex optical field. The micro-comb produced a large number of comb lines via a CMOS-compatible nonlinear MRR that greatly increases the processing bandwidth. By programming and shaping the comb lines according to the calculated tap weights, we successfully demonstrate a fractional differentiator with tunable fractional orders ranging from 0.15 to 0.9. We characterized the RF amplitude and phase response, obtaining an operation bandwidth of ~15.49 GHz. We perform system demonstrations of the real-time fractional differentiator for Gaussian pulse input signals, obtaining good agreement with theory and verifying that this approach is an effective way to implement high-speed reconfigurable fractional differentiators featuring high processing bandwidths and reconfigurability, for future ultra-high-speed RF systems.

## V. Acknowledgements

This work was supported by the Australian Research Council Discovery Projects Program (No. DP150104327). RM acknowledges support by Natural Sciences and Engineering Research Council of Canada (NSERC) through the Strategic, Discovery and Acceleration Grants Schemes, by the MESI PSR-SIIRI Initiative in Quebec, and by the Canada Research Chair Program. He also acknowledges additional support by the Government of the Russian Federation through the ITMO Fellowship and Professorship Program (grant 074-U 01) and by the 1000 Talents Sichuan Program in China.

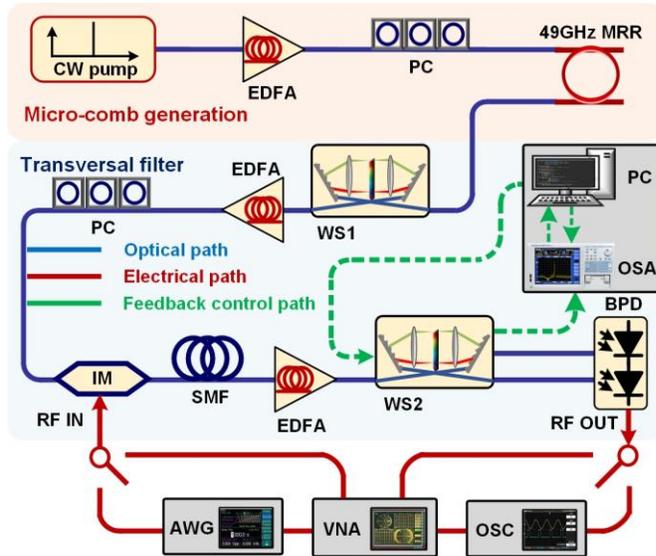

Fig. 1. Experimental setup of a fractional differentiator based on a microcomb. EDFA: erbium-doped fiber amplifier. PC: polarization controller. MRR: micro-ring resonator. WS: WaveShaper. IM: Intensity modulator. SMF: single mode fiber. BPD: Balanced photodetector, AWG: arbitrary waveform generator, VNA: Vector Network analyzer.

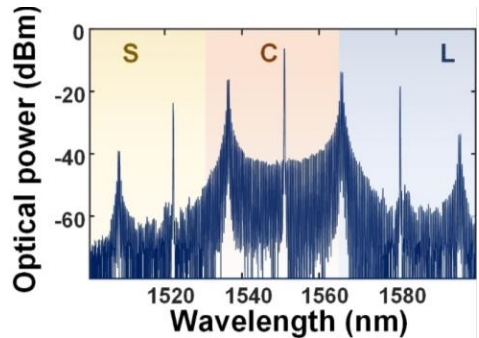

Fig. 2. The optical spectrum of the soliton crystal microcomb.

.

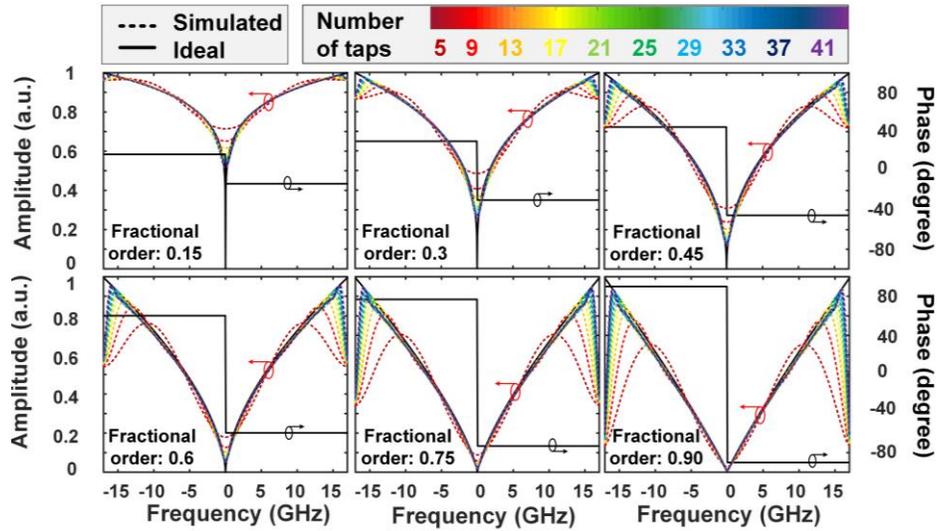

Fig. 3. Simulated transfer function of different fractional differentiation orders with varying number of taps.

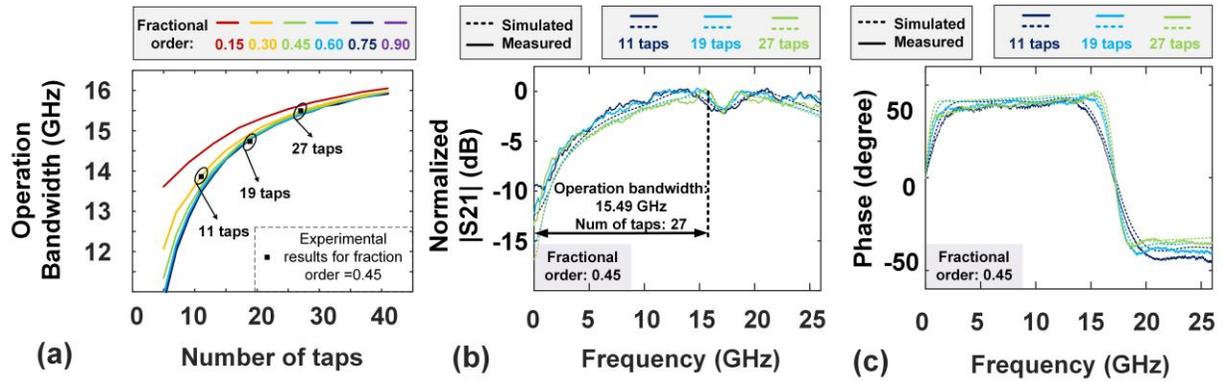

Fig. 4. (a) Extracted relationship between the number of taps and operation bandwidth. (b, c) Experimentally demonstrated fractional differentiator with varying number of taps.

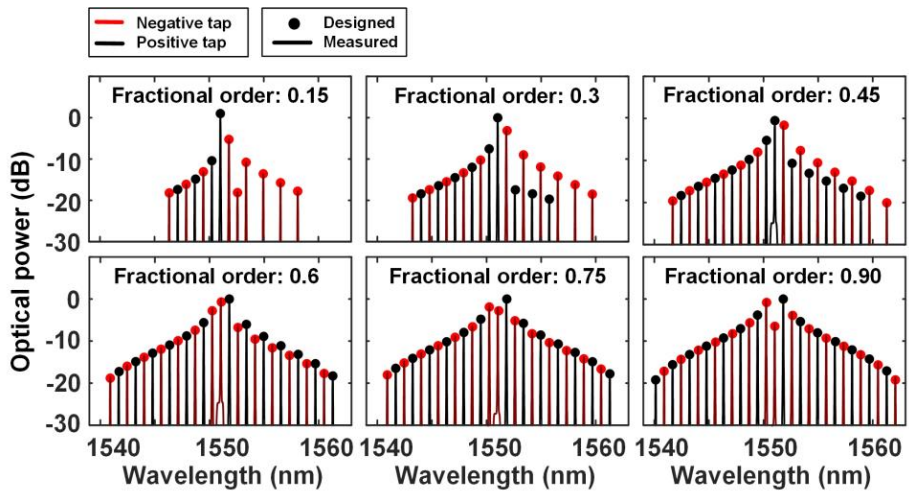

Fig. 5. Optical spectra of the shaped microcomb for different fractional orders.

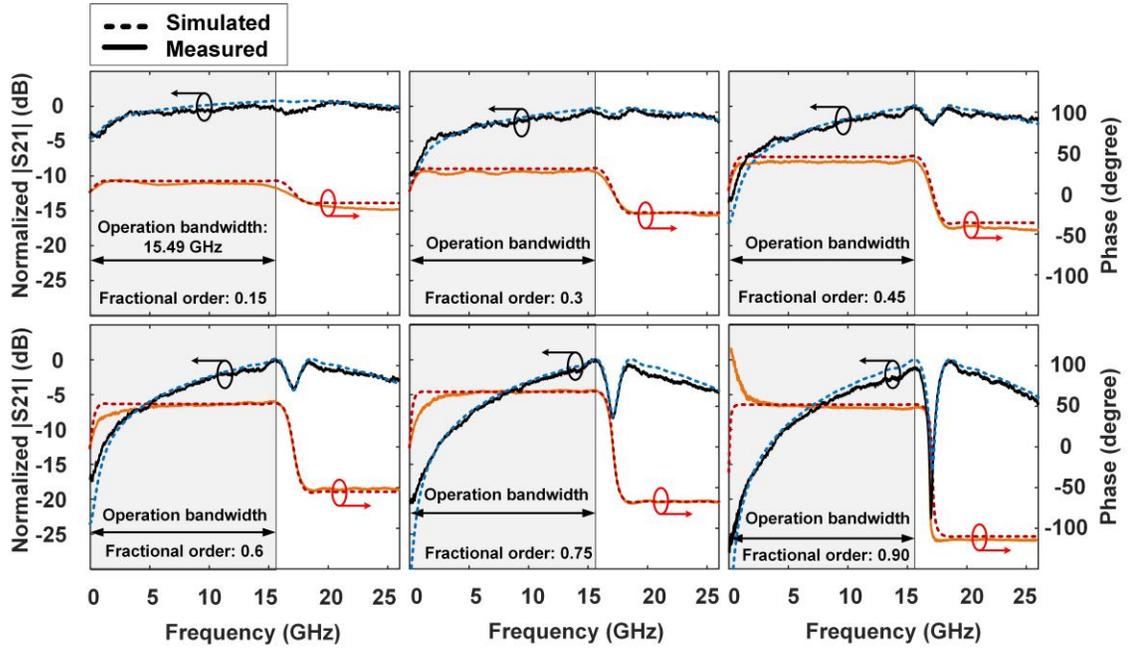

Fig. 6. Simulated and measured the transmission response of the fractional differentiator at different orders ranging from 0.15 to 0.90.

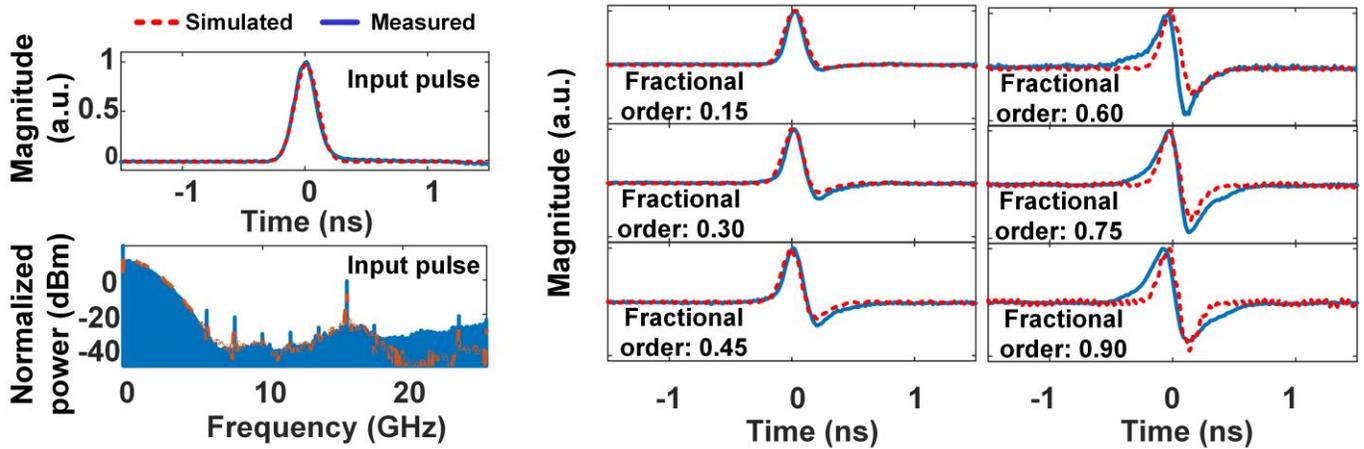

Fig. 7. Simulated and measured RF Gaussian pulse and the output temporal intensity waveform after the fractional differentiator.